\documentclass{iopart}
\usepackage{graphicx}
\begin{document}

\title{Dynamics of a magnetized Bianchi I universe with vacuum energy}
\author{Emma J King and Peter Coles} \address{Centre for Astronomy \&
  Particle Theory, School of Physics \& Astronomy, University of
  Nottingham, NG7 2RD, UK} \ead{ppxejk@nottingham.ac.uk}

\begin{abstract}
  We make use of a flat, axisymmetric Bianchi I metric to investigate
  the effects of a magnetic field upon the dynamics of the universe
  for the case in which the accompanying fluid is
  a cosmological constant and derive two exact solutions to the
  dynamical equations for this situation.  We examine the behaviour of the
  scale factor perpendicular and parallel to the field lines, $A(t)$
  and $W(t)$ respectively, and find the expected behaviour.  The field
  has the strongest effect when $A(t)$ is small, decelerating collapse
  perpendicular to the field lines, due to magnetic pressure, and
  accelerating collapse along the field lines, due to magnetic
  tension, while the vacuum energy dominates at late time, driving
  accelerated expansion.
\end{abstract}

\section{Introduction}
\label{sec:intro}

The Bianchi models, which describe homogeneous but anisotropic
spacetimes, have recently undergone a resurgence in popularity,
motivated in part by attempts to explain small but significant
anisotropies in the cosmic microwave background (CMB), as seen in data
from satellites such as WMAP (e.g. \cite{jbegh05}).  Such models can
admit large-scale magnetic fields which, if of sufficient strength,
may be able to affect the overall expansion rate of the universe.  The
axisymmetric Bianchi I model is appropriate for examining the case of
a universe which is permeated by a large scale, homogeneous magnetic
field.  This model has been studied in the literature for the case
when the field is accompanied by a perfect fluid with $0 \leq \gamma
\leq 1$, where $\gamma$ is the barotropic index of the fluid.  The
case where the accompanying fluid is a cosmological constant,
$\Lambda$, which has $\gamma = -1$, has, however, historically been
ignored.  This is primarily because these models were under
investigation in the late 1960s, when a cosmological constant was not
generally considered to be a relavent or realistic constituent of the
universe. Since then, we have learned that the universe is currently
dominated by dark energy, which may be in the form of a cosmological
constant (\cite{r98,p99,k03,r06}), and that the universe underwent
exponential expansion for a short period, known as inflation, in its
early history, during which time the inflaton acted as an effective
cosmological constant \cite{g81} (for a recent review see
\cite{t02}).  Hence models containing a cosmological constant are now
of significant interest.

The observed level of isotropy in the CMB places tight constraints
upon the strength of any Hubble scale magnetic field, and the upper
limit on such a field is found to be of the order of $10^{-9} G$
\cite{bfs97}. It is important to remember, however, that these and
other similar limits quote the adiabatically expanded, present-day
equivalent value of the field strength.  The strength of a magnetic
field is proportional to the density of its field lines and so,
assuming the field is frozen into the plasma that expands with the
Hubble flow,
\begin{equation}
B \propto \frac{1}{a^2(t)},
\end{equation}
where $B$ is the magnetic field strength and $a(t)$ is the scale
factor of the universe.  Thus although equivalent to a very weak
field at the present time, these limits equate to a much stronger
field at very early times, when the scale factor is small.  In
particular, the exponential rate of increase of the scale factor
during inflation means that a significant field could exist during
the early stages of inflation before being diluted away and becoming
consistent with later constraints.  An understanding of the effect
of a magnetic field upon the dynamics of the universe in the
presence of a cosmological constant may, therefore, have
implications for models of the universe which contain a magnetic
field generated before or during inflation (e.g.
\cite{tw88,r92,ds93,d93,lp04,am05}).

In this paper we use the axisymmetric Bianchi I model to look at the
effects of a large-scale magnetic field upon the dynamics of the
universe in the presence of a cosmological constant.  In section
\ref{sec:Bianchi} we outline the axisymmetric Bianchi I model and
give the fundamental dynamical equations that describe its evolution
in the presence of both a perfect, barotropic fluid and a large
scale magnetic field.  In section \ref{sec:GR_existsolns} we briefly
review the main findings which arise from existing solutions to
these equations from the literature, with particular emphasis upon
the general behaviour and effects of the magnetic field in these
solutions.  In section \ref{sec:GR_newsolns} we derive two exact
solutions for a magnetized universe with vacuum energy and in
section \ref{sec:GR_cases} we examine the results of those solutions
for various specific cases and in certain limits.  Finally, in
section \ref{sec:GR_disc} we discuss our findings.

\section{Axisymmetric Bianchi I model}
\label{sec:Bianchi}

The assumption that the universe is both homogeneous and isotropic
restricts the spacetime metric to the Freidmann--Robertson--Walker
(FRW) model.  Relaxation of the condition of isotropy allows extra
degrees of freedom and the Bianchi models are the resulting set of
homogeneous but anisotropic spacetimes, first classified by Bianchi in
1918.  A good review of anisotropic spacetimes, which gives details of
the Bianchi classifications, can be found in \cite{ll75}; see also
\cite{e06}.

The Bianchi I model describes a universe in which the scale factor is
not restricted to be the same in each direction, as it is in the FRW
model. We wish to consider a flat universe in which the anisotropy is
introduced by a large-scale homogeneous magnetic field.  Such a
magnetic field will impose a single preferred direction in space,
along the field lines, and so the anisotropy of the spacetime will be
axisymmetric.  Hence we use the flat, axisymmetric Bianchi I metric,
\begin{equation}
\label{eq:bt1metric} ds^2=-dt^2+A^2(t)[dx^2+dy^2]+W^2(t)dz^2,
\end{equation}
as the basis for describing such a universe.  This metric is
homogeneous, geometrically flat and has two equivalent
\emph{transverse} directions, $x$ and $y$, and one different
\emph{longitudinal} direction, $z$, along which we assume the magnetic
field is orientated ($\bi{B}=B_z$ only).  $A(t)$ is then the scale
factor in the transverse direction, perpendicular to the magnetic
field, while $W(t)$ is the scale factor in the longitudinal direction,
along the field lines.  In the limit where $A(t) = W(t) = a(t)$,
(\ref{eq:bt1metric}) reduces to the flat FRW metric.

We assume that there is no electric field, that the magnetic field is
orientated along the $z$ axis and that the universe is filled with a
perfect, barotropic fluid with equation of state
\begin{equation}
 \label{eq:perfectfluid}
 p_f = \gamma \rho_f,
\end{equation}
where $\rho$ is the energy density, $p$ is the pressure, and a
subscript '$f$' indicates the fluid.  The combined stress--energy
tensor for both the magnetic field and the fluid is then given by
\begin{equation}
 \label{eq:stressenergy}
 T^{\mu}_{\ \nu} = {\rm diag}~\left( -\rho_f-\rho_B, p_f+\rho_B, p_f+\rho_B,
 p_f-\rho_B \right),
\end{equation}
where a subscript '$B$' indicates the magnetic field.  The Einstein field
equation,
\begin{equation}
 G_{\mu\nu} = 8 \pi T_{\mu\nu},
\end{equation}
then leads to three equations which describe the dynamics of the
spacetime \cite{j69,d65},
\begin{equation}
 \label{eq:D1}
 16\pi\rho_B =
 \frac{\ddot{A}}{A}+\left(\frac{\dot{A}}{A}\right)^2-\frac{\dot{A}}{A}\frac{\dot{W}}{W}-\frac{\ddot{W}}{W},
\end{equation}
\begin{equation}
 \label{eq:D2}
 16\pi\rho_f =
 -\frac{\ddot{A}}{A}+\left(\frac{\dot{A}}{A}\right)^2+5\frac{\dot{A}}{A}\frac{\dot{W}}{W}+\frac{\ddot{W}}{W}
\end{equation}
and
\begin{equation}
 \label{eq:D3}
 16\pi\gamma\rho_f =
 -3\frac{\ddot{A}}{A}-\left(\frac{\dot{A}}{A}\right)^2-\frac{\dot{A}}{A}\frac{\dot{W}}{W}-\frac{\ddot{W}}{W}.
\end{equation}
Assuming conservation of energy,
\begin{equation}
 \label{eq:energycons}
 T^{\mu}_{~\nu;\mu} = 0,
\end{equation}
where a semicolon indicates a covariant derivative, for the fluid and
the magnetic field leads to two equations \cite{j69},
\begin{equation}
 \label{eq:mattercons}
 \rho_f = \frac{\mu}{8
 \pi}\frac{1}{\left(A^2W\right)^{\left(1+\gamma\right)}}
\end{equation}
and
\begin{equation}
 \label{eq:fluxcons}
 \rho_B = \frac{\beta}{8 \pi}\frac{1}{A^4}
\end{equation}
respectively, where $\mu$ and $\beta$ are positive constants.

Equations (\ref{eq:D1})--(\ref{eq:D3}), (\ref{eq:mattercons}) and
(\ref{eq:fluxcons}) describe the dynamics of a Bianchi I universe
which contains a magnetic field of energy density $\rho_B$ and a
perfect, barotropic fluid of energy density $\rho_f$ and equation of
state given by (\ref{eq:perfectfluid}).  It is possible to solve these
equations to find $A(t)$ and $W(t)$ which then describe the expansion
of the universe in the transverse and longitudinal directions
respectively.

\section{General properties}
\label{sec:GR_existsolns}

The equations from section \ref{sec:Bianchi} have been used in the
literature to investigate the effects of a large-scale uniform
magnetic field on the dynamics of the universe when the accompanying
fluid has $0 \leq \gamma \leq 1$ (e.g. \cite{j69,d65,t67}). A wide
range of scenarios has been investigated; here we restrict ourselves
to considering those in a flat Bianchi I cosmology.  For example,
\cite{t67} contains analytic solutions for the case of dust ($\gamma
= 0$) and radiation ($\gamma = 1/3$), while \cite{j69} also finds
analytic solutions for $\gamma = 1$ and for $1/3 \leq \gamma \leq
1$.  In addition, they find a special case of the $\gamma = 1$
solution in which $\mu = 0$ and hence there is no fluid---the
\emph{pure magnetic} case. For general properties of these
solutions, see \cite{skm03}.

Both \cite{j69,t67} find that the asymptotic nature of
these solutions varies depending on the values of the various
constants.  They can collapse isotropically or anisotropically to a
\emph{point} singularity, collapse in the longitudinal direction thus
forming a \emph{pancake} singularity, or collapse in the transverse
direction thus forming a \emph{cigar} singularity.  In general the
existing solutions follow several `rules of thumb' that describe the
effect of the magnetic field upon the dynamical evolution of the
universe, which are given in \cite{t67}.  These are:
\begin{enumerate}
\item If $\rho_B \ll \rho_f$ and the model is not approaching a
  singularity then the magnetic field has negligible effect upon the
  dynamics.
\item In general, the magnetic field accelerates expansion (or
  decelerates collapse) in the transverse direction owing to magnetic
  pressure, which resists compression of the field lines, and
  decelerates expansion (or accelerates collapse) in the longitudinal
  direction owing to tension in the field lines.
\item Near a singularity of infinite density ($\rho_f \to \infty$), if
  $\rho_B/\rho_f$ (which is always positive) $\to 0$ as the
  singularity is approached then the magnetic field has negligible
  effect upon the dynamics.
\item Near a singularity of infinite density, if $\rho_B/\rho_f$ does
  not approach 0 then one of two cases occur.  Either (a)
  $\rho_B/\rho_f \to {\rm const}$, in which case the fluid and
  magnetic field jointly determine the dynamics, or (b) the magnetic
  field causes rapid expansion in the transverse direction (see point
  (ii)), and this change in the dynamics causes $\rho_B/\rho_f$ to
  approach zero.
\end{enumerate}

It is worth commenting on the magnetic Bianchi models with dust or
radiation. In these cases there is a significant effect on the
evolution of the shear, in that it falls off much more slowly than
in a model  without the magnetic field. This is why the cosmic
microwave background gives a strong constraint \cite{bfs97} on
homogeneous magnetic fields \cite{jb97}.

\section{Solutions with vacuum energy}
\label{sec:GR_newsolns}

The existing solutions, as discussed in section
\ref{sec:GR_existsolns}, concentrate on cases where the fluid has $0
\leq \gamma \leq 1$.  In this section we present two solutions to the
dynamical equations which describe a magnetised Bianchi I universe
containing a cosmological constant, $\gamma = -1$, one of which is
physically interesting. These belong to a general class of solutions
that has been discussed in the literature before, but from a
different perspective \cite{lb97} to that discussed here; see also
\cite{skm03} for these solutions in a different form.

When $\gamma = -1$, $\rho_f = \rho_{\Lambda}$, the energy density of
the cosmological constant, or vacuum energy.  Equations
(\ref{eq:D1})--(\ref{eq:D3}) then become
\begin{equation}
 \label{eq:D1-1}
 16\pi\rho_B =
 \frac{\ddot{A}}{A}+\left(\frac{\dot{A}}{A}\right)^2-\frac{\dot{A}}{A}\frac{\dot{W}}{W}-\frac{\ddot{W}}{W},
\end{equation}
\begin{equation}
 \label{eq:D2-1}
 16\pi\rho_{\Lambda} =
 -\frac{\ddot{A}}{A}+\left(\frac{\dot{A}}{A}\right)^2+5\frac{\dot{A}}{A}\frac{\dot{W}}{W}+\frac{\ddot{W}}{W}
\end{equation}
and
\begin{equation}
 \label{eq:D3-1}
 -16\pi\rho_{\Lambda} =
 -3\frac{\ddot{A}}{A}-\left(\frac{\dot{A}}{A}\right)^2-\frac{\dot{A}}{A}\frac{\dot{W}}{W}-\frac{\ddot{W}}{W}
\end{equation}
respectively, and from (\ref{eq:mattercons}), the vacuum energy is
given by
\begin{equation}
  \rho_{\Lambda} = \frac{\mu}{8 \pi}.
\end{equation}

The left-hand sides of (\ref{eq:D2-1}) and (\ref{eq:D3-1}) are now
equal and opposite.  There are two ways in which the right-hand sides
of these equations can be equal and opposite, and these give rise to
our two solutions.  The easiest route is to assume that all
derivatives of $A$ are equal to zero, which leads us to conclude that
$A=constant$.  Substituting this into (\ref{eq:D3-1}) gives
\begin{equation}
 \label{eq:Wunphys}
 \frac{\ddot{W}}{W}=2\mu.
\end{equation}
Substituting this result into (12) and (13) and adding these equations
immediately shows that either $\rho_B$, $\rho_\Lambda$ or $\mu$ must
all be identically zero (the empty universe case), or at least one
must be negative, in which case this solution is unphysical.

An alternative solution can be found by equating (\ref{eq:D2-1}) and
(\ref{eq:D3-1}) which, after some cancellation, reduces to
\begin{equation}
 \label{eq:W-1_2}
 \frac{\dot{W}}{W}=\frac{\ddot{A}}{\dot{A}},
\end{equation}
from which we can immediately see that
\begin{equation}
 \label{eq:Wtk}
 W=k\dot{A},
\end{equation}
where $k$ is a constant of integration.  Substituting
(\ref{eq:mattercons}) and (\ref{eq:fluxcons}) back into
(\ref{eq:D1-1}) and (\ref{eq:D3-1}), adding these equations and
substituting (\ref{eq:Wt}) into the result leads to
\begin{equation}
 \label{eq:Wdiffk}
 \frac{\beta}{A^4}+\mu = k^2\left(\left(\frac{W}{A}\right)^2+2\frac{WW'}{A}\right),
\end{equation}
where a prime denotes differentiation with respect to $A$.  The
constant $k$ can be removed from this equation by scaling both $\beta$
and $\mu$ by a factor of $k^2$.  Hence the value of $k$ does not
affect the form of the solution and so we are free to set it to unity,
giving
\begin{equation}
 \label{eq:Wt}
 W=\dot{A}
\end{equation}
and
\begin{equation}
 \label{eq:Wdiff}
 \frac{\beta}{A^4}+\mu = \left(\frac{W}{A}\right)^2+2\frac{WW'}{A}.
\end{equation}
Equation (\ref{eq:Wdiff}) can now be solved to find $W(A)$, giving
\begin{equation}
 \label{eq:WA}
 W=\frac{1}{3A}\left( 3 \mu A^4 -9\beta +9cA\right)^{1/2},
\end{equation} where $c$ is a constant of integration, which, from
(\ref{eq:Wt}), leads to
\begin{equation}
 \label{eq:t}
 t = \int{3A\left(3\mu A^4 - 9\beta +9cA\right)^{-1/2} dA}.
\end{equation}

Equations (\ref{eq:Wt}) and (\ref{eq:t}) form our second solution,
which is of more physical interest than the first. Unfortunately, it
is not easy to integrate (\ref{eq:t}) analytically, making it
difficult to find an explicit expression for $t(A)$ and hence $A(t)$
and $W(t)$.  Instead, in the next section we consider the form of
$\dot{A}$ in different limits and use this to gain an insight into the
form of $A(t)$ and $W(t)$ and hence into the expansion history of the
universe.

\section{Specific cases}
\label{sec:GR_cases}

From (\ref{eq:WA}) we can make deductions about the form of $W(A)$,
and hence of $A(t)$, depending upon the value of the constants $\mu$,
$\beta$ and $c$ and in different limits.  In this section we consider
first the empty universe case, $\mu, \beta = 0$, before examining the
effect of the magnetic field and vacuum energy independently.
Finally, we look at the combined case, $\mu, \beta > 0$.

\subsection{Empty universe ($\mu, \beta =0$)}
\label{sec:GR_cases_empty}

When $\mu, \beta = 0$, (\ref{eq:WA}) reduces to
\begin{equation}
 \label{eq:WemptyaC}
 W = \left(\frac{c}{A}\right)^{1/2},
\end{equation} giving
\begin{equation}
 \label{eq:Aempty}
 A=c^{1/3}\left(\frac{3}{2}t\right)^{2/3}.
\end{equation}
Substituting this back into (\ref{eq:WemptyaC}) gives
\begin{equation}
 \label{eq:WemptyC}
 W=\left(\frac{3}{2c}t\right)^{-1/3}.
\end{equation} 
Coordinate freedom in the metric given in equation
(\ref{eq:bt1metric}) allows us to scale away the factors of $c$ in
both $A(t)$ and $W(t)$, leaving
\begin{equation}
 \label{eq:Aempty}
 A=\left(\frac{3}{2}t\right)^{2/3}
\end{equation}
and
\begin{equation}
 \label{eq:Wempty}
 W=\left(\frac{3}{2}t\right)^{-1/3},
\end{equation} respectively.

\begin{figure}
  \centering $\begin{array}{cc}
    \includegraphics[width=0.5\textwidth]{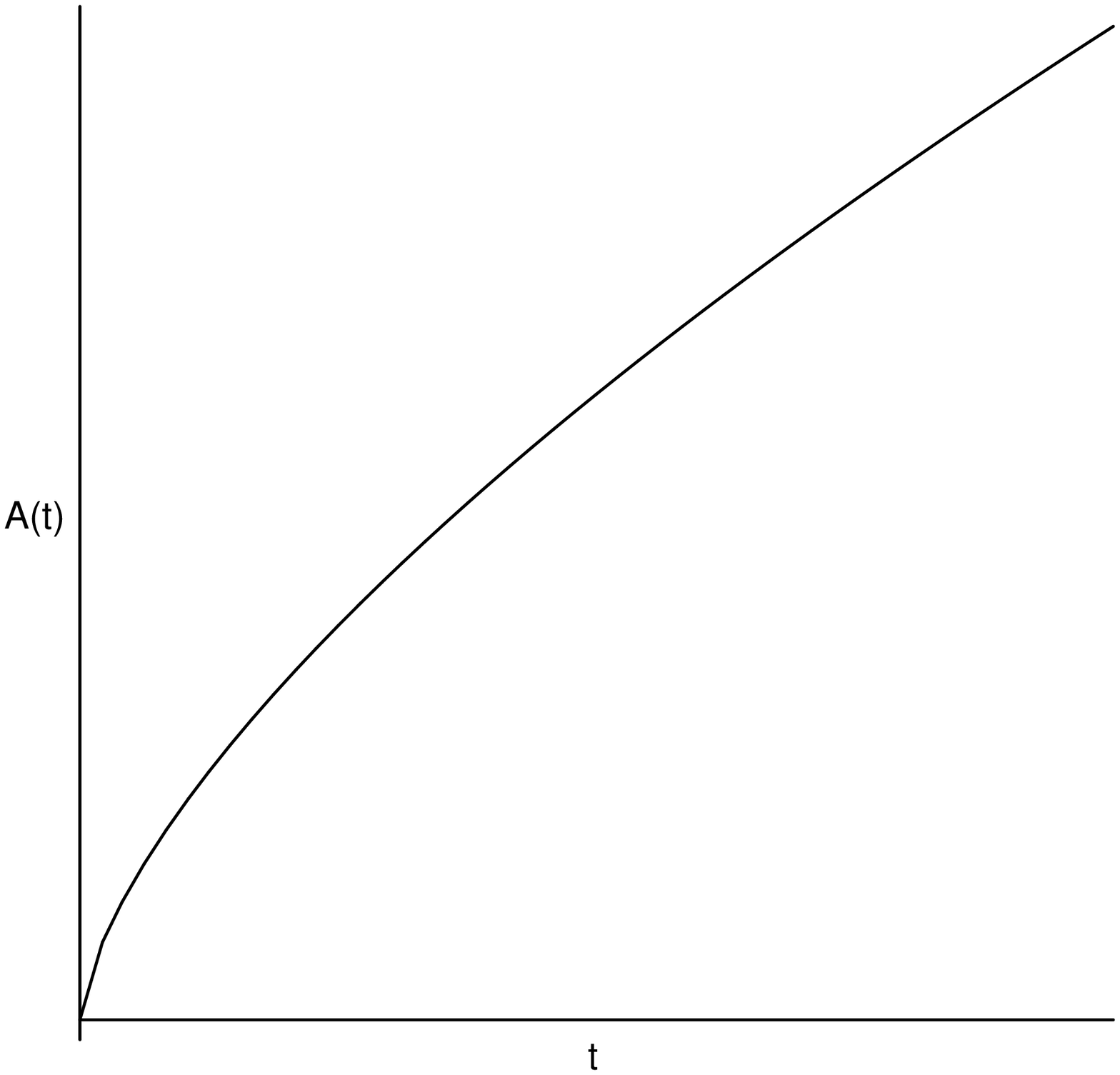} &
    \includegraphics[width=0.5\textwidth]{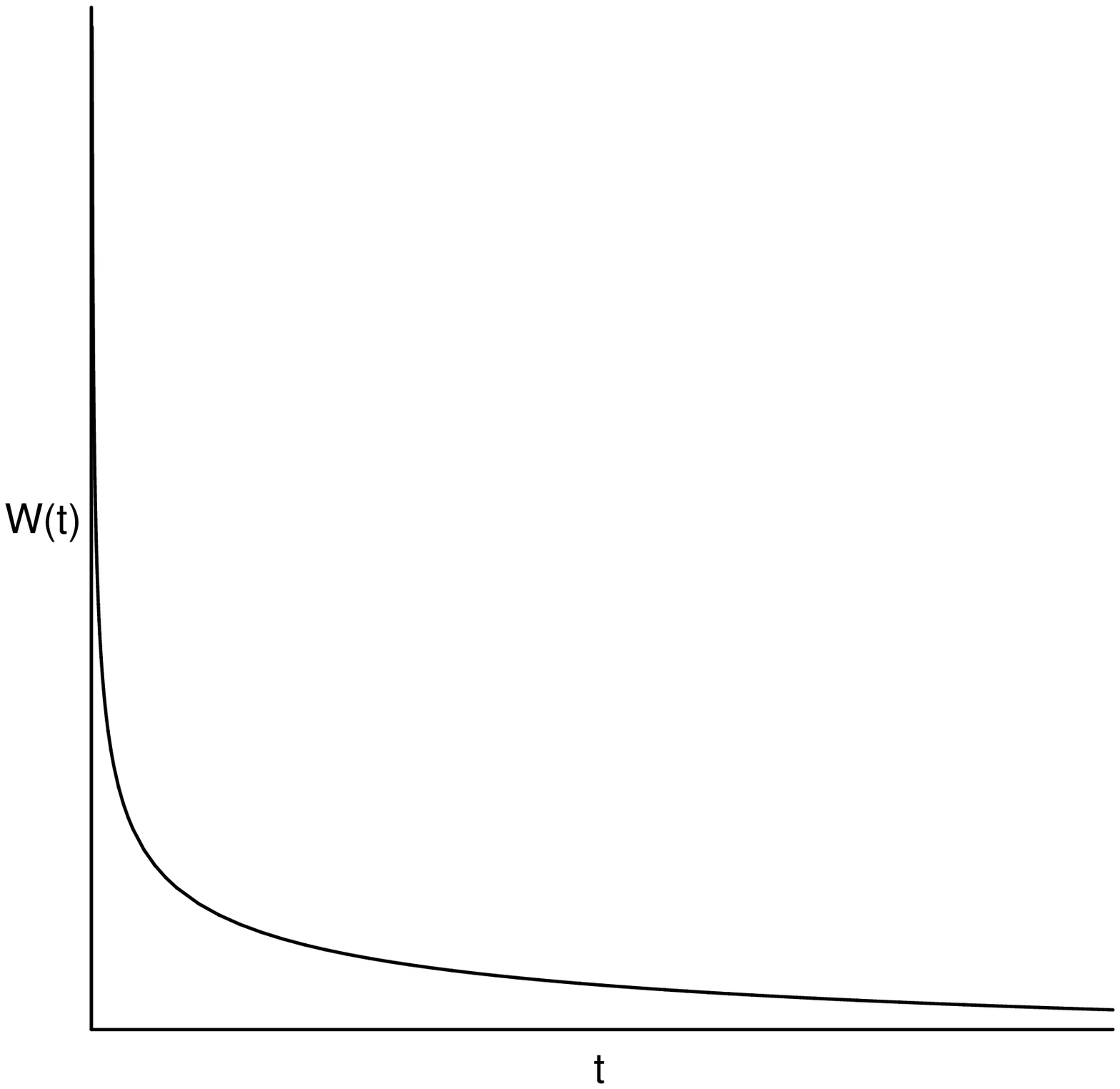} \\
    (a) & (b)
  \end{array}$
  \caption[$(a)$ $A(t)$ and $(b)$ $W(t)$ for the empty universe
  case]{\label{fig:AWempty}Evolution of $(a)$ $A(t)$ and $(b)$ $W(t)$
    for the empty universe case.}
\end{figure}

Equations (\ref{eq:Aempty}) and (\ref{eq:Wempty}) describe the
evolution of the empty universe in the transverse and longitudinal
directions respectively, as illustrated in figure \ref{fig:AWempty}.
In this case the universe emerges from an infinitely long cigar
singularity ($A=0$, $W=\infty$) at $t=0$.  As $t$ increases, $W$
decreases asymptotically to zero while $A$ continues to increase,
leading to another singularity at $t=\infty$.  This final singularity
is an infinitely wide pancake ($A = \infty$, $W = 0$).

Reassuringly, this solution is one of two well-known possible
solutions to the empty axisymmetric Bianchi I model, also known as the
Kasner universe (see, e.g.,  \cite{ll75,skm03,k21}).  The other possible
solution was the special, empty universe case of our first solution.
We can see directly from (\ref{eq:D1})--(\ref{eq:D3}) that $A =
constant$, $W = constant$ is also a solution when $\mu, \beta = 0$.
This is the case of a static empty universe and it is this which
reduces to the static solution to the empty FRW model when $A(t) =
W(t)$.

\subsection{Pure magnetic ($\mu = 0$)}
\label{sec:GR_cases_pureB}

When $\mu=0$ we have the pure magnetic case, with a magnetic field but
no vacuum energy.  In this case (\ref{eq:WA}) reduces to
\begin{equation}
 \label{eq:WBa}
 W=\frac{\left(cA-\beta\right)^{1/2}}{A},
\end{equation}
from which we can see that $c$ must be a positive constant for this
solution to be physical.  Hence
\begin{equation}
 \label{eq:tB}
 t = \frac{2}{3}\left(2\beta +cA\right)\sqrt{cA-\beta}.
\end{equation}
The only real solution to this equation is given by
\begin{equation}
 \label{eq:AB}
 A= \frac{1}{c}\left(\beta+\left(\frac{1}{2}x-\frac{2\beta}{x}\right)^2
\right),
\end{equation}
where
\begin{equation}
 \label{eq:x}
 x=\left(6tc^2+2\sqrt{16\beta^3+9c^4t^2}\right)^{1/3}.
\end{equation} Substituting this back into (\ref{eq:WBa}) gives
\begin{equation}
\label{eq:WB}
W=\frac{c\left(\frac{1}{2}x-\frac{2\beta}{x}\right)^{1/2}}
{\beta+\left(\frac{1}{2}x-\frac{2\beta}{x}\right)^2}.
\end{equation}
Both $A$ and $W$ again have constant factors which can be scaled away
due to coordinate freedom in (\ref{eq:bt1metric}); however the factors
of $c$ included in $x$ cannot be removed in the same fashion.

\begin{figure}
  \centering $\begin{array}{cc}
    \includegraphics[width=0.5\textwidth]{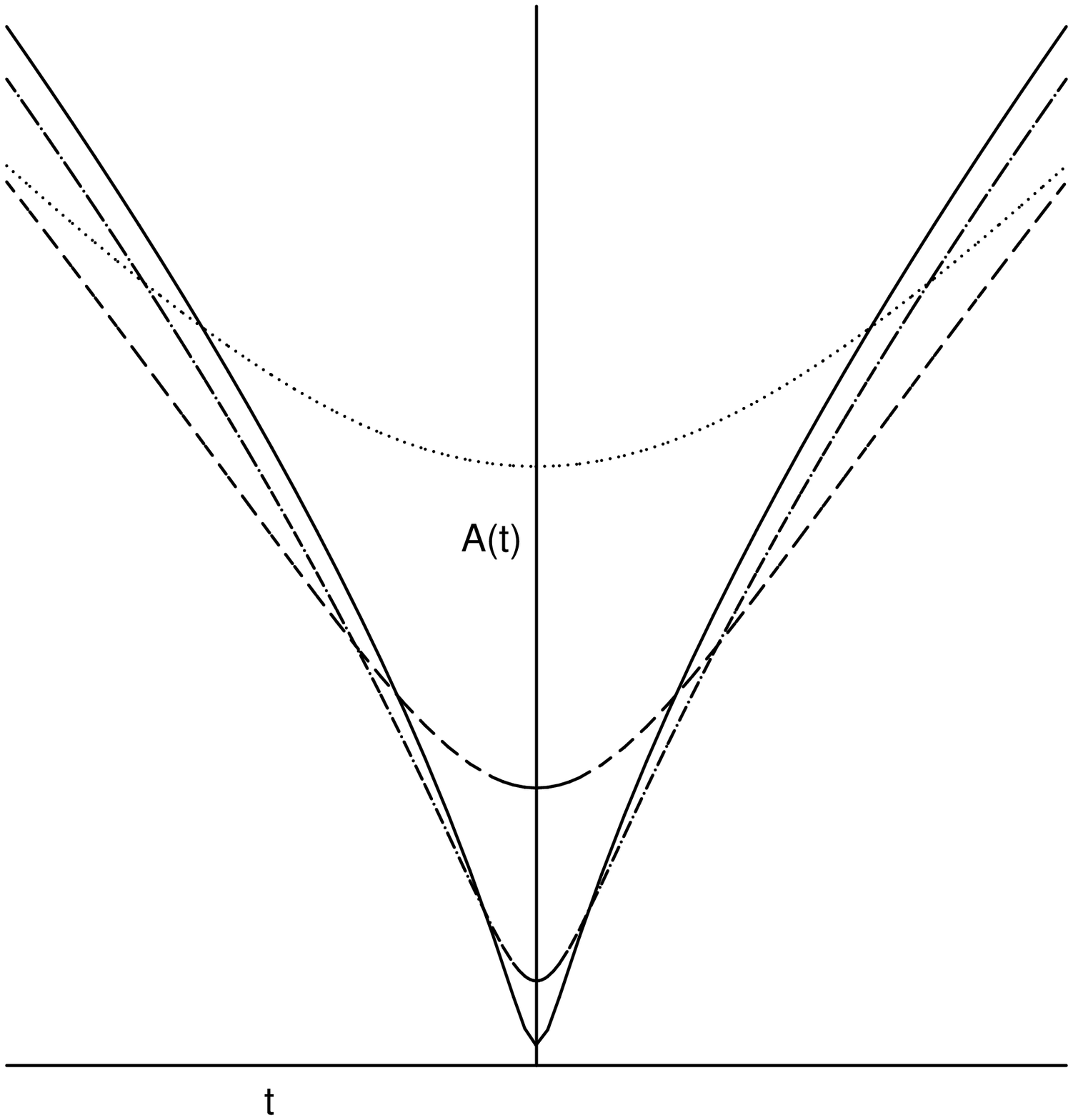} &
    \includegraphics[width=0.5\textwidth]{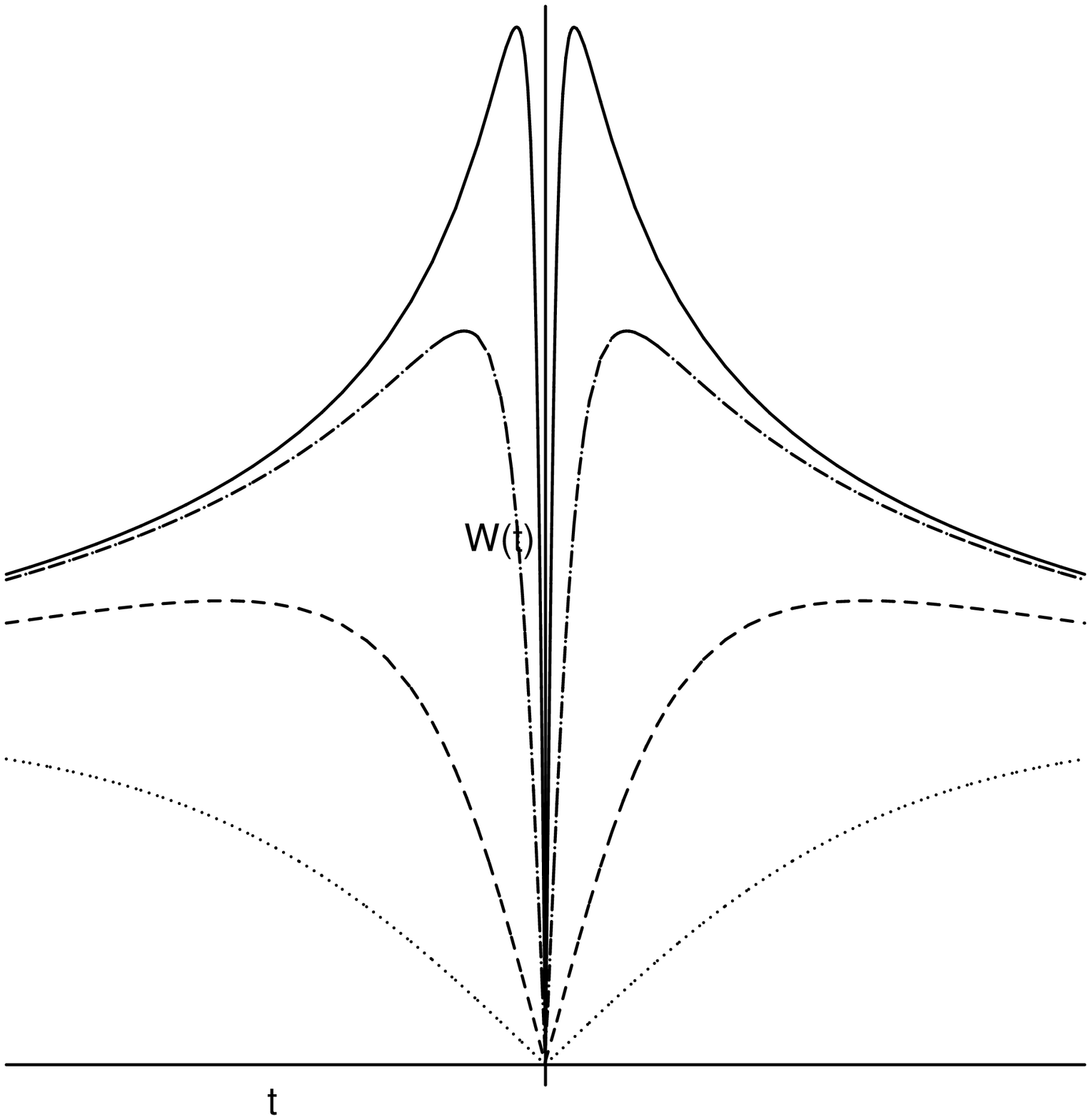} \\
    (a) & (b)
  \end{array}$
  \caption[$(a)$ $A(t)$ and $(b)$ $W(t)$ for the pure magnetic
  case]{\label{fig:AWB}Evolution of $(a)$ $A(t)$ and $(b)$ $W(t)$ for
    the pure magnetic case plotted for $c=1$ and varying values of
    $\beta$, as follows: solid, $\beta = 1$; dot-dashed, $\beta = 2$;
    dashed, $\beta = 5$; dotted, $\beta = 10$.}
\end{figure}

The form of $A(t)$ and $W(t)$ in this case can be seen in figure
\ref{fig:AWB},
where they are both plotted for $c=1$ and varying values of $\beta$ in
order to illustrate the effect of $\beta$ upon the expansion.  The
effect of increasing $c$ is to increase the initial rate of expansion
of both $A$ and $W$, hence increasing the gradient of both graphs at
small $t$.

Although we are able to solve for $A(t)$ and $W(t)$ exactly, it
is instructive to consider independently the limits of large and small
$A$ in (\ref{eq:WBa}) in order to understand the behaviour of $A$ and
$W$ at early and late times.

For small $A$ ($A \sim \beta/c$), the magnetic field has a profound
effect upon the dynamics of the universe.  As we approach the initial
singularity it prevents collapse in the transverse direction, due to
magnetic pressure, and accelerates collapse in the longitudinal
direction, due to tension in the field lines.  This causes a bounce at
a pancake singularity ($A>0$, $W=0$) at $t=0$. The value of $A$ at
the initial singularity is determined by $\beta$ as expected, since a
stronger magnetic field will provide a greater pressure resisting the
collapse.  The strength of the field also affects the tension in the
field lines; hence the maximum value of $W$ and the time at which this
is reached also depends upon $\beta$.

For large $A$ ($A \gg \beta/c$), the magnetic field becomes negligible
and (\ref{eq:WBa}) again reduces to the empty universe case.  Hence,
far from the initial singularity the magnetic field has no effect, as
would be expected for large values of $A$, since $\rho_B$ is very
small here.  We therefore have a second pancake singularity at
$t=\infty$ (mirrored at $t=-\infty$), exactly as for the empty
universe.

We can also see from these limits that the value of $c$ influences the
point at which the magnetic regime moves into the empty universe
regime, with this happening at smaller values of $A$ (and hence
earlier times) for larger values of $c$.

It is interesting to note that the behaviour of $A$ and $W$ in this
case is qualitatively the same as for the axisymmetric pure magnetic
case in \cite{j69}, which was found from the solutions for $A(t)$ and
$W(t)$ when $\gamma = 1$.  This solution is therefore probably the
same as theirs for this case, as would be expected, but arrived at via
a different route.

\subsection{Pure $\Lambda$ ($\beta = 0$)}
\label{sec:GR_cases_pureL}

When $\beta=0$ we have the pure $\Lambda$ case, with vacuum energy but
no magnetic field.  In this case, (\ref{eq:WA}) reduces to
\begin{equation}
 \label{eq:WLa}
 W=\frac{\left(3\mu A^4 + 9cA\right)^{1/2}}{3A},
\end{equation}
which gives
\begin{equation}
 \label{eq:tL}
 t = \int{\frac{3A}{\left(3\mu A^4 + 9cA\right)^{1/2}}dA}.
\end{equation}
This expression is not easy to integrate using analytical techniques.  It
is clear, however, that, in the denominator of (\ref{eq:tL}),
the $A$ term will dominate in the limit of small $A$, while the $A^4$
term will dominate in the limit of large $A$.  These two regimes meet
when
\begin{equation}
 A  =  \left(3c/\mu\right)^{1/3}.
\end{equation}
Hence we can consider the behaviour of $A(t)$ and $W(t)$ in these two
limits.

\begin{figure}
  \centering $\begin{array}{cc}
    \includegraphics[width=0.5\textwidth]{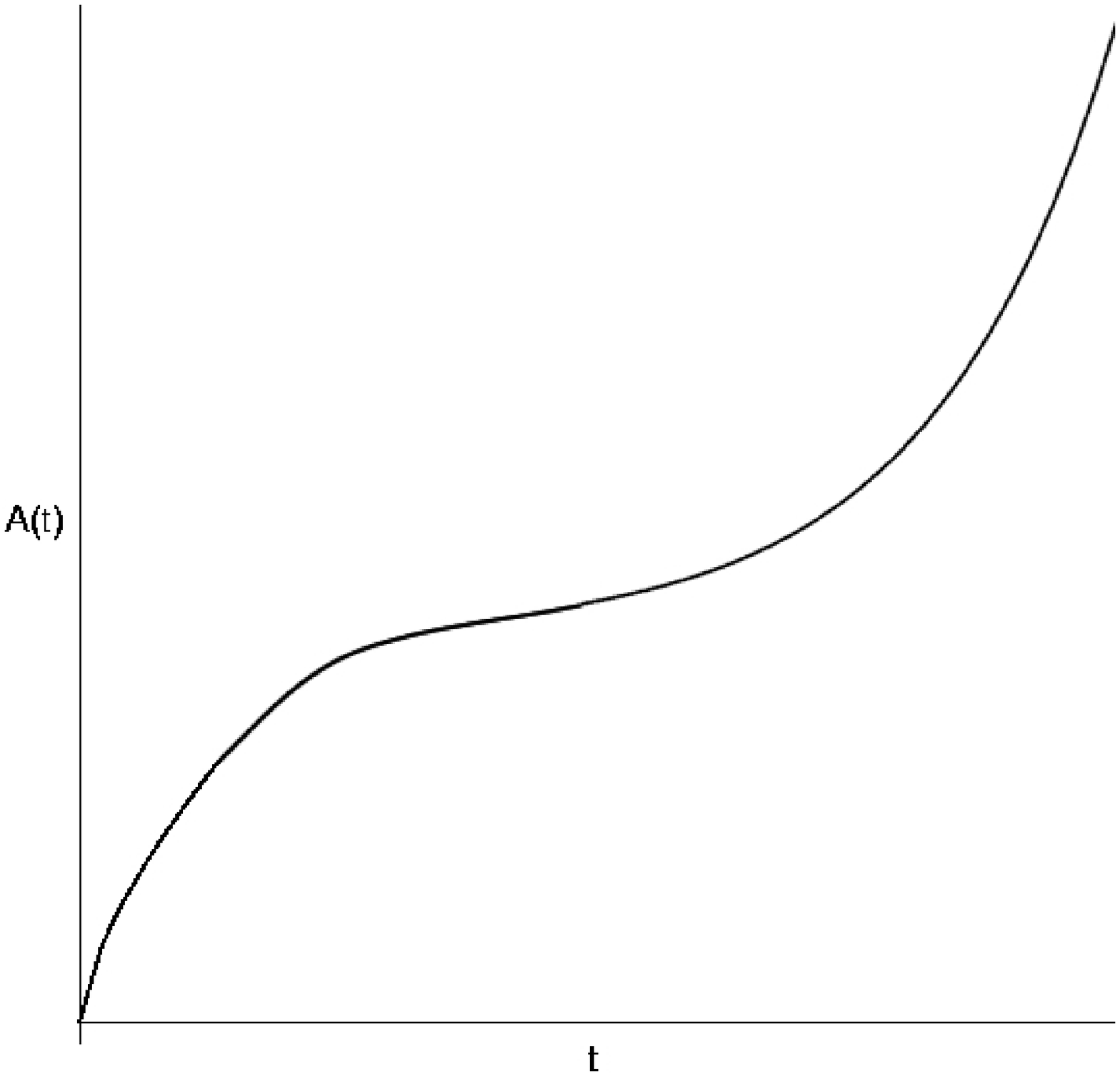} &
    \includegraphics[width=0.5\textwidth]{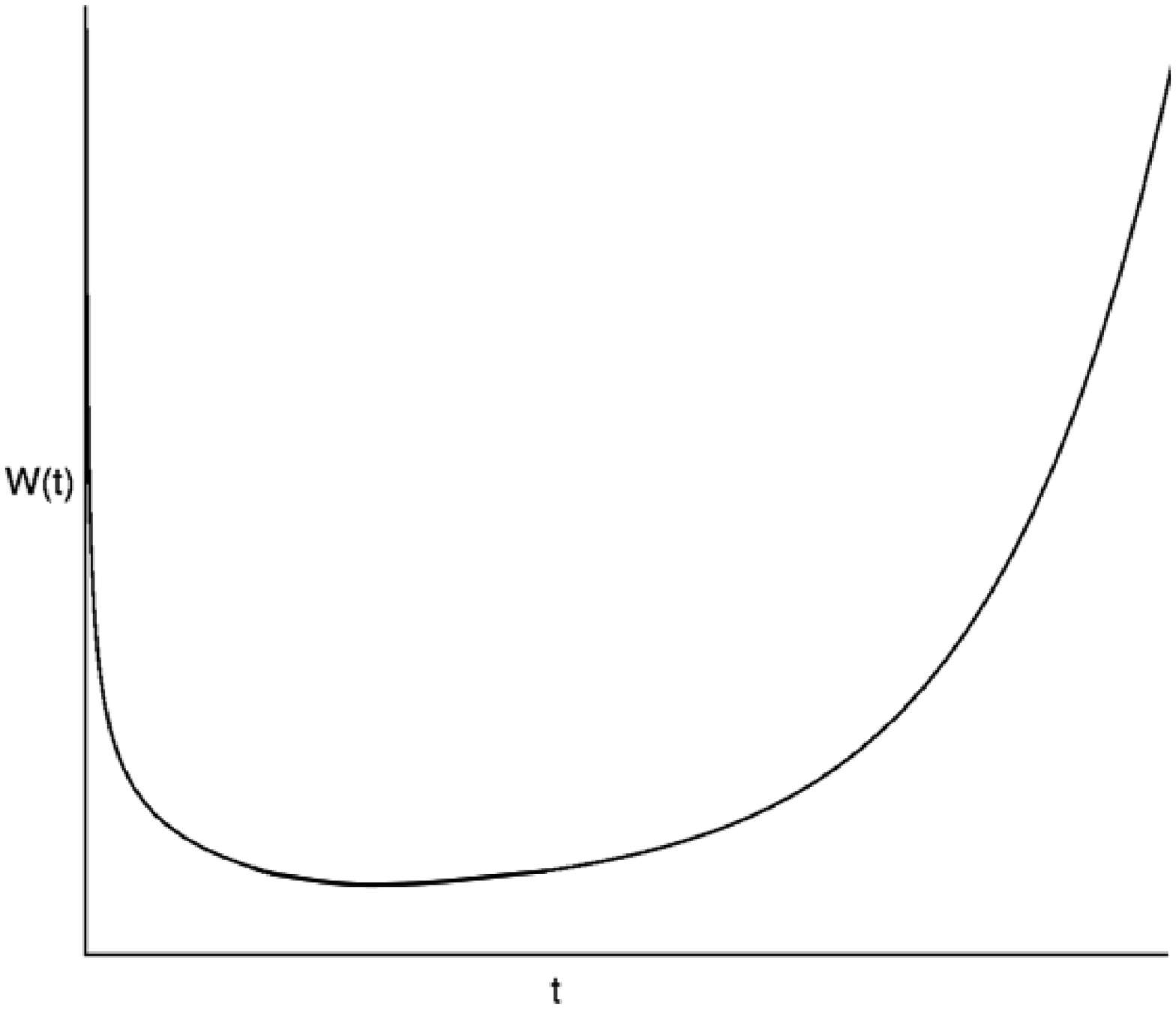} \\
    (a) & (b)
  \end{array}$
  \caption[$(a)$ $A(t)$ and $(b)$ $W(t)$ for the pure $\Lambda$
  case]{\label{fig:AWL}Evolution of $(a)$ $A(t)$ and $(b)$ $W(t)$ for
    the pure $\Lambda$ case.  The point at which the solution turns
    from the empty universe to cosmological constant-driven expansion
    is controlled by the values of $\mu$ and $c$.}
\end{figure}

In the limit of small $A$ ($A \ll (3c/\mu)^{1/3}$), (\ref{eq:WLa})
reduces to the empty universe case, as described in section
\ref{sec:GR_cases_empty}. In the limit of large $A$ ($A \gg
(3c/\mu)^{1/3}$) it reduces to
\begin{equation}
 \label{eq:WLal}
 W=\sqrt{\frac{\mu}{3}}A,
\end{equation}
which gives
\begin{equation}
 \label{eq:ALl}
 A=\rme^{\sqrt{\mu/3}t}.
\end{equation}
Substituting this back into (\ref{eq:WLal}) then gives
\begin{equation}
 \label{eq:WLl}
 W=\sqrt{\frac{\mu}{3}}\rme^{\sqrt{\mu/3}t},
\end{equation}
and so both $A(t)$ and $W(t)$ increase exponentially in this limit (see
figure \ref{fig:AWL}).

This analysis suggests that the universe emerges from an infinitely
long cigar singularity at $t=0$, as in the empty universe, with $A$
increasing steadily and $W$ decreasing rapidly.  $W$ then reaches a
minimum value, after which it begins to increase.  At large $t$, both
$A$ and $W$ increase exponentially, leading to an infinitely large,
ever expanding universe without a final singularity.  This exponential
increase in $W$ and $A$ at late times can be understood as the effect
of the vacuum energy driving an accelerated expansion.  The
accelerated expansion begins when $A \sim (3c/\mu)^{1/3}$; hence
the larger the value of $\mu$, and the higher the energy density of
the cosmological constant, the earlier its effect is seen, as
expected.  The higher the value of $c$, on the other hand, the longer
the empty universe regime persists.

\subsection{Combined case}
\label{sec:GR_cases_comb}

\begin{figure}
  \centering $\begin{array}{cc}
    \includegraphics[width=0.5\textwidth]{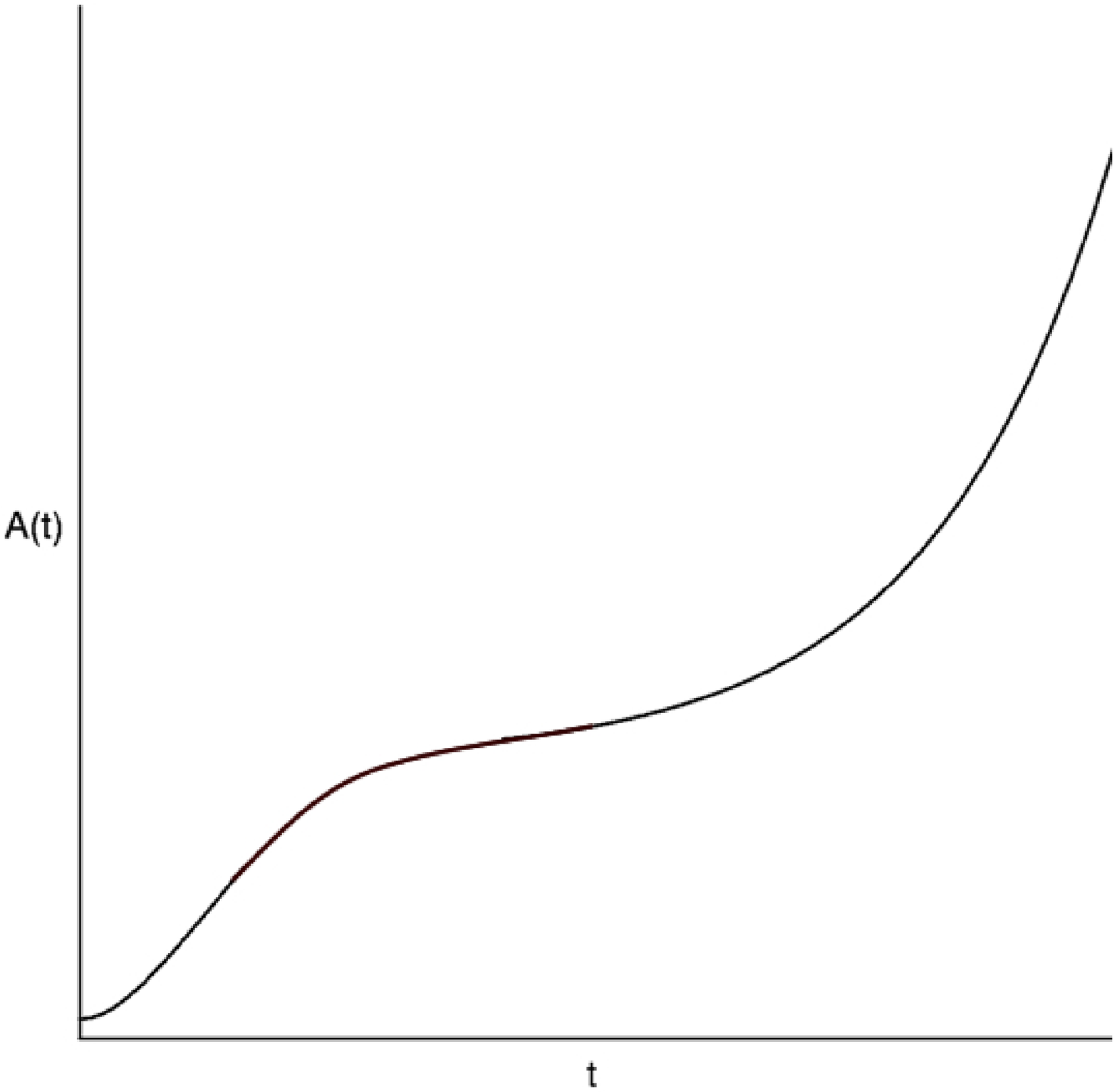} &
    \includegraphics[width=0.5\textwidth]{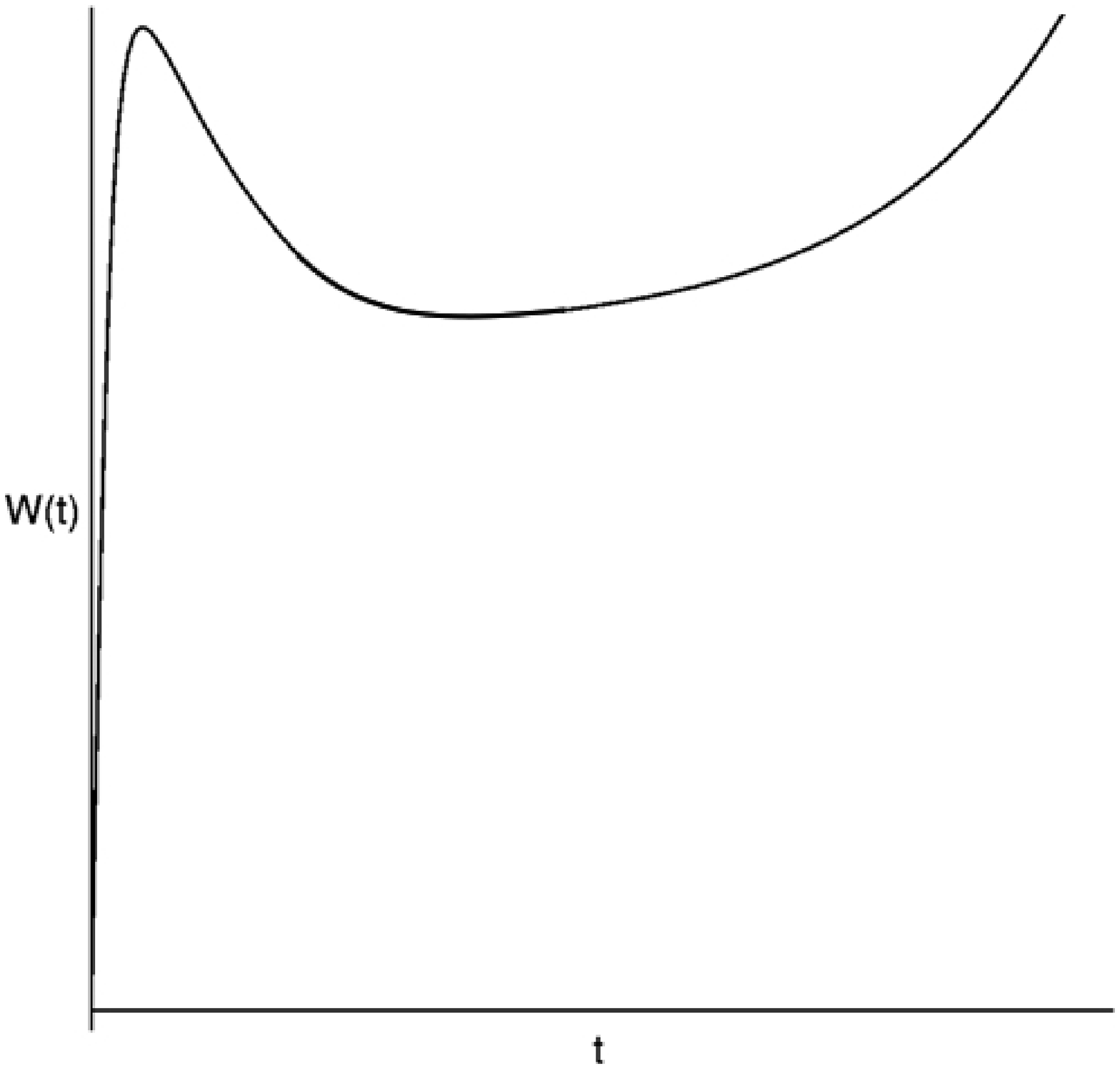} \\
    (a) & (b)
  \end{array}$
  \caption[$(a)$ $A(t)$ and $(b)$ $W(t)$ for the combined
  case]{\label{fig:AWC}Evolution of $(a)$ $A(t)$ and $(b)$ $W(t)$ for
    the combined magnetic and cosmological constant case.}
\end{figure}

For the combined case, $\mu, \beta > 0$, which contains both a
magnetic field and vacuum energy, we cannot integrate $t(A)$ so must
again look at the limits where different terms dominate in
(\ref{eq:WA}).  Although mathematically it is possible to consider two
regimes, domination by the $-9\beta + 9cA$ term at small $A$ and by the
$3 \mu A^4$ term at large $A$, it is again instructive to split the
early regime into two.  Hence we consider three possible regimes, as
follows:
\begin{enumerate}
\item $A \sim \beta/c$ and $A \ll (3c/\mu)^{1/3}$---The square root in
  (\ref{eq:WA}) is dominated by $9cA-9\beta$.  The magnetic field
  dominates as in the early pure magnetic case.
\item $A \gg \beta/c $ and $A \ll (3c/\mu)^{1/3}$---The square root in
  (\ref{eq:WA}) is dominated by the $9cA$ term.  Neither magnetic field
  nor cosmological constant affect the evolution, which proceeds as
  for an empty universe.
\item $A \gg \beta/c $ and $A \gg (3c/\mu)^{1/3}$---The square root in
  (\ref{eq:WA}) is dominated by the $3\mu A^4$ term.  The vacuum energy
  dominates the expansion as in the late pure $\Lambda$ case.
\end{enumerate}
Hence the universe has a bounce at a finite pancake singularity ($A >
0$, $W=0$) at $t=0$, from which $A$ and $W$ both increase.  Depending
upon the values of $\mu$, $\beta$ and $c$ it may then enter a regime
where $W$ decreases while $A$ continues to increase, as in the empty
universe case.  Finally, the vacuum energy comes to dominate causing
both $A$ and $W$ to increase exponentially.  The behaviour of $A(t)$
and $W(t)$ for this case is illustrated in figure \ref{fig:AWC}.

This behaviour is unsurprising; we expect the influence of the
magnetic field to be greater at smaller values of $A$, since the
magnetic field strength is proportional to $1/A^2$, and the
cosmological constant to dominate at later times, since its energy
density remains constant as the universe expands.  Hence as $A$
increases, the effect of the magnetic field is diminished and the
cosmological constant comes to dominate once the magnetic field has
been diluted away.

The time at which these regimes cross depends upon the values of
$\mu$, $\beta$ and $c$.  The higher the value of $\beta$, the stronger
the magnetic field and the longer it will continue to influence the
dynamics of the universe.  Similarly, the higher the value of $\mu$,
the higher the energy density of the cosmological constant and the
earlier it will come to dominate.  If both $\mu$ and $\beta$ are
sufficiently large compared to $c$, the magnetic regime will move
directly into the $\Lambda$ regime and there will be no empty universe
regime.  We leave the magnetic regime at $A \gg \beta/c$ and enter the
$\Lambda$ regime at $A \gg (3c/\mu)^{1/3}$; thus there is no empty
universe regime if these two limits cross, i.e. if
\begin{equation}
 \label{eq:limitx}
 \mu \beta^3 \gg 3c^4.
\end{equation}

\section{Discussion}
\label{sec:GR_disc}

We have obtained two exact solutions to the dynamical equations for
an axisymmetric Bianchi I universe containing a magnetic field and a
perfect, barotropic fluid for the previously unconsidered case of a
cosmological constant with $\gamma = -1$.  The first of these
solutions is unphysical, leading to imaginary values of the
transverse scale factor, $A(t)$, if $\beta > 0$.  The second
solution is more physically interesting, and we have used it to
investigate the effect of a magnetic field on the dynamics of the
spacetime in the presence of vacuum energy.

The magnetic field has the strongest effect upon the dynamics at early
times, when $A$ is small.  It tends to decelerate collapse in the
transverse direction and accelerate collapse in the longitudinal
direction as the initial singularity is approached.  The vacuum
energy, on the other hand, comes to dominate the expansion at late
time and causes accelerated expansion in both the transverse and
longitudinal directions.

Additionally, we find that the shape of the initial singularity
depends upon the presence or absence of the magnetic field, with the
field transforming it from a cigar to a pancake singularity, and
that the existence of a final singularity depends upon the presence
or absence of the cosmological constant, which prevents the final
singularity from occurring.  It should be noted that all the
singularities we find are physical, as can be seen from the Ricci
curvature scalar, which becomes infinite in each case.

These results are unsurprising; we expect the effect of the magnetic
field to be greatest when $\rho_B$ is highest, which occurs at low
values of $A$, and that of the cosmological constant to be greatest at
late times, since its energy density is not diluted by cosmological
expansion.  We also expect magnetic pressure to decelerate collapse
(or accelerate expansion) in the transverse direction, while magnetic
tension accelerates collapse (or decelerates expansion) in the
longitudinal direction, and the vacuum energy to drive accelerated
expansion.  Our results clearly follow the `rules of thumb' for the
effect of the magnetic field given by \cite{t67}, which were outlined
in section \ref{sec:GR_existsolns}.

Finally, we find that the constant $c$ affects the initial rate of
expansion, which is greater for higher values of $c$.  In addition the
empty universe regime will become dominant at earlier times, and
persist for longer, for larger values of $c$.

Since our model is homogenous and spatially flat, magnetic tension
effects arise from the negative magnetic pressure along the field
lines. In more general magnetic cosmologies there is also the
possibility that tension arises from inhomogeneities. The implications
of these other effects have been discussed elsewhere in the literature
\cite{mt00,tm00,ts01,ts06}, but they are of a physically different
origin and should not be confused with those we discuss here.

Our model cannot be compared directly to the observable universe,
which contains significant quantities of dust and radiation as well
as dark energy and which is not currently observed to have any
significant anisotropy which would indicate the presence of a large
scale magnetic field.  It is possible, however, that the universe
may have contained a significant strength magnetic field at early
times and so these results may have implications for the evolution
of the universe during inflation, when the inflaton acts as an
effective cosmological constant.  It is clear from our results that
the effect of the exponential expansion at late times is to
drastically dilute the magnetic field, $\rho_B \to 0$, and to
isotropise the initially highly anisotropic universe.  Similar
processes may have operated in the early universe and hence any
magnetic field created before or during inflation could have had a
significant impact upon the dynamics of the universe before being
diluted away in the same fashion.

\section*{Acknowledgments}

We thank Sigbjorn Hervik, John Barrow, Christos Tsagas, Roy Maartens
and Bob Kirshner for helpful comments on an earlier version of this
paper.

\section*{References}

\bibliographystyle{unsrt}
\bibliography{KingColes}

\end{document}